# Ultrawide Band Gap β-Ga₂O₃ Nanomechanical Resonators with Spatially Visualized Multimode Motion

Xu-Qian Zheng, Jaesung Lee, Subrina Rafique, Lu Han,

Christian A. Zorman, Hongping Zhao, Philip X.-L. Feng[*]

*Department of Electrical Engineering & Computer Science, Case School of Engineering,*

*Case Western Reserve University, 10900 Euclid Avenue, Cleveland, OH 44106, USA*

### *Abstract*

**Beta gallium oxide (β-Ga₂O₃) is an emerging ultrawide band gap (4.5 eV–4.9 eV) semiconductor with attractive properties for future power electronics, optoelectronics, and sensors for detecting gases and ultraviolet radiation. β-Ga₂O₃ thin films made by various methods are being actively studied toward such devices. Here, we report on the experimental demonstration of single-crystal β-Ga₂O₃ nanomechanical resonators using β-Ga₂O₃ nanoflakes grown via low-pressure chemical vapor deposition (LPCVD). By investigating β-Ga₂O₃ circular drumhead structures, we demonstrate multimode nanoresonators up to the 6ᵗʰ mode in high and very high frequency (HF / VHF) bands, and also realize spatial mapping and visualization of the multimode motion. These measurements reveal a Young's modulus of $E_Y$ = 261 GPa and anisotropic biaxial built-in tension of 37.5 MPa and 107.5 MPa. We find that thermal annealing can considerably improve the resonance characteristics, including ~40% upshift in frequency and ~90% enhancement in quality ($Q$) factor. This study lays a foundation for future exploration and development of mechanically coupled and tunable β-Ga₂O₃ electronic, optoelectronic, and physical sensing devices.**

---

[*]Corresponding Author.  Email:  philip.feng@case.edu

Wide band gap (WBG) semiconductors, such as gallium nitride (GaN) and silicon carbide (SiC) have been widely employed in power electronics, high-temperature and harsh-environment electronics and transducers, as well as ultraviolet (UV) photonics,[1,2,3,4,5] thanks to their advanced material properties such as high critical field strength ($\mathscr{E}_{br,GaN} \approx 3.3$ MV cm$^{-1}$, $\mathscr{E}_{br,4H-SiC} \approx 2.5$ MV cm$^{-1}$) and UV photon absorption (band gap $E_{g,GaN} \approx 3.4$ eV, $E_{g,SiC} \approx 2.3-3.3$ eV). These WBG crystals also possess outstanding mechanical properties such as very high Young's modulus ($E_{Y,GaN} \approx 300$ GPa, $E_{Y,SiC} \approx 400$ GPa), exceptional thermal properties (*e.g.*, thermal conductivity $\kappa_{GaN} = 130$ W m$^{-1}$ K$^{-1}$, and $\kappa_{SiC} = 360-490$ W m$^{-1}$ K$^{-1}$, at ~300K), outstanding thermal tolerance (maintaining mechanical properties up to 1300 °C) and chemical stability.[6] These characteristics, combined with readily available microfabrication techniques, make these WBG crystals technologically critical materials for advanced micro/nanoelectromechanical systems (M/NEMS) beyond the conventional counterparts based upon silicon (Si). By exploiting these advantageous properties, great progresses have been made in SiC and GaN M/NEMS,[6,7,8,9] for sensing and detection of physical quantities, and low-power signal processing at radio frequencies, especially for high-temperature and harsh-environment scenarios.

As an emerging ultra-wide band gap (UWBG) semiconductor, β-Ga$_2$O$_3$ has recently attracted increasing interest due to its UWBG ($E_{g,\beta-Ga_2O_3} \approx 4.5-4.9$ eV) that is significantly wider than those of GaN and SiC.[10,11] It offers very high critical field strength ($\mathscr{E}_{br,\beta-Ga_2O_3} = 8$ MV cm$^{-1}$ predicted, and $\mathscr{E}_{br,\beta-Ga_2O_3} = 3.8$ MV cm$^{-1}$ measured) and electron mobility up to $\mu_n = 300$ cm$^2$ V$^{-1}$·s$^{-1}$ at room temperature.[12,13,14,15,16] These make β-Ga$_2$O$_3$ a promising contender for future generation power devices.[15,17,18] Beyond electrical properties, β-Ga$_2$O$_3$ also exhibits excellent mechanical strength (Young's modulus $E_{Y,\beta-Ga_2O_3} = 232$ GPa) along with extraordinary chemical and thermal (melting point at 1820 °C) stability.[19,20] The excellent ensemble of attributes in β-Ga$_2$O$_3$ makes it suitable for new UWBG M/NEMS beyond SiC and GaN M/NEMS. Further, β-Ga$_2$O$_3$ is sensitive to solar-blind UV light,[21,22] and offers reversible response to oxidation and reduction gases,[23] enabling UV light and gas sensing applications. Importantly, in contrast to other demonstrated WBG M/NEMS materials including SiC, diamond, and GaN, bulk β-Ga$_2$O$_3$ crystals can be made by homoepitaxy (thus preventing threading dislocations) and other more cost-effective melting growth methods, such as Czochralski (CZ),[24,25] floating zone (FZ),[26,27,28] and edge-defined film-fed growth (EFG)[29] techniques. The variety of synthesis methods have already led to realization of various β-Ga$_2$O$_3$ nanostructures, including nanowires,[30] nanobelts,[31] nanorods,[32] and nanosheets.[33] Given these exciting and promising properties, in parallel to currently active pursuit of electronic devices, it is intriguing to explore β-Ga$_2$O$_3$ as a new structural material for M/NEMS; and it is important to gain knowledge and understanding of its mechanical properties down to the nanoscale. Towards these objectives, here we perform the first experiments on studying and exploiting mechanical properties of β-Ga$_2$O$_3$ nanoflakes by realizing the first robust β-Ga$_2$O$_3$ nanomechanical devices, from material synthesis to device fabrication and characterization. Employing all-dry-transfer techniques, we fabricate suspended β-Ga$_2$O$_3$ diaphragms based on β-Ga$_2$O$_3$ nanoflakes [34] synthesized by low-pressure chemical vapor deposition (LPCVD).[35,36] Using an ultrasensitive laser optical interferometry measurement system, we detect the undriven thermomechanical motion of these nanoresonators, from which we examine thermal annealing effects and resolve the elastic properties of the β-Ga$_2$O$_3$ nanoflakes. Further, we analyze the frequency scaling laws of multimode β-Ga$_2$O$_3$ nanoresonators and discover the anisotropic stress distribution in the β-Ga$_2$O$_3$ diaphragms, by comparing finite element modeling (FEM) results with measured multimode resonances and their mode shapes.

β-Ga$_2$O$_3$ is the most stable polymorph of Ga$_2$O$_3$ which has a monoclinic crystalline structure (Figure 1a).[37] The construction of β-Ga$_2$O$_3$ suspended nanomechanical resonators starts with the synthesis of β-Ga$_2$O$_3$ nanoflakes. Using a LPCVD method,[34] we grow β-Ga$_2$O$_3$ nanostructures on a 3C-SiC epi-layer on Si substrate at a growth temperature of 950 °C for 1.5 hours, using high purity Ga pellets and O$_2$ (Ar as carrier gas) as source materials (Figure S1 in Supporting Information (SI)). The as-grown nanoflakes have widths of ~2–30 μm and thicknesses of ~20–140 nm (Figure 1b). We use an all-dry-transfer method to fabricate the suspended β-Ga$_2$O$_3$ diaphragms (Figure 1c, Figure S2 in SI). The as-grown nanoflakes are picked up by thermal release tape which is then stamped onto a substrate with arrays of pre-defined microtrenches that are ~290 nm deep (Figure 1c). By heating the structure up to 90 °C and gently lifting the thermal release tape, the β-Ga$_2$O$_3$ nanoflakes remain on the substrate, forming circular suspended diaphragms with diameters of ~3.2 μm and ~5.2 μm (set by the sizes of the pre-patterned circular microtrenches). To examine the crystal quality of β-Ga$_2$O$_3$ after device fabrication, we first measure Raman spectra of the transferred flakes and find Raman modes at ~143 cm$^{-1}$, ~168 cm$^{-1}$, ~199 cm$^{-1}$, ~346 cm$^{-1}$, ~416 cm$^{-1}$, ~476 cm$^{-1}$, ~653 cm$^{-1}$, and ~768 cm$^{-1}$, excellently matching the calibrated Raman modes of bulk β-Ga₂O₃; this verifies the high crystal quality of the suspended β-Ga₂O₃ diaphragms (Figure S3 in SI).[38]

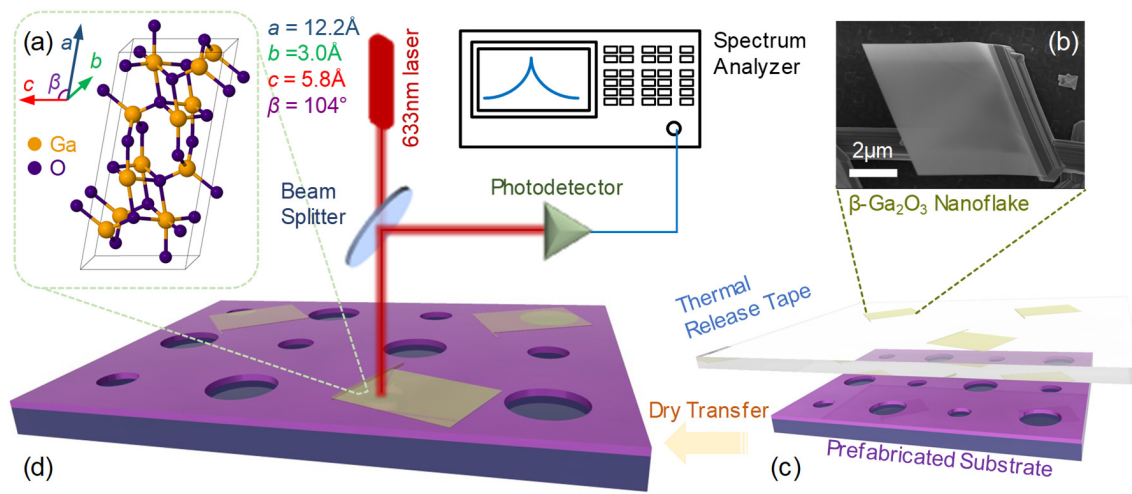

**Figure 1.** Schematic representation of experimental demonstration of β-Ga₂O₃ nanomechanical resonators. (a) β-Ga₂O₃ crystal structure. (b) SEM image of an as-grown β-Ga₂O₃ nanoflake extruding from a β-Ga₂O₃ nanorod grown on the 3C-SiC substrate. (c) Illustration of all-dry transfer of β-Ga₂O₃ nanoflakes by using thermal release tape as a stamp to create suspended β-Ga₂O₃ devices on pre-defined microtrenches and arrays. (d) Illustration of fabricated β-Ga₂O₃ resonators under study by using the scanning laser interferometry motion detection and spatial mapping system.

We then conduct optical resonance measurements on the β-Ga₂O₃ circular drumhead diaphragms: a He−Ne (633 nm) laser is focused upon the suspended β-Ga₂O₃ device to interferometrically detect the out-of-plane motion of the diaphragm (Figure 1d). The laser scanning interferometry detection system has been engineered to achieve:[39] (i) ultrahigh displacement sensitivity (~4 fm Hz$^{-1/2}$) in reading out flexural vibrations down to the undriven, Brownian thermomechanical motion, and (ii) submicron-scale spatial resolution in its scanning spectromicroscopy mode, thus capable of spatially mapping and vividly visualizing the mode shapes of the multimode resonances.

We carefully investigate the multimode resonances of these devices, by conducting thermal annealing at 250 °C for 1.5 hours and calibrating resonance characteristics both before and after annealing. Fundamental mode resonance spectra of a typical device with diameter $d$~3.2 μm (Figure 2a) before and after annealing are shown in Figure 2b and Figure 2c, respectively. Fitting the spectra to a damped simple harmonic resonator model, we find the thermal annealing has led to a resonance frequency ($f$) upshift from 65.50 MHz to 74.90 MHz and quality ($Q$) factor enhancement from 420 to 566. Measured fundamental mode $f$s and $Q$s of 10 devices (Figure 2d-f) show an average $f$ increase of ~40 % and a $Q$ enhancement of ~90 %. The results suggest that thermal annealing (with same parameters) may be comparatively more effective for these β-Ga$_2$O$_3$ nanoresonators than for their molybdenum disulfide (MoS$_2$) or hexagonal boron nitride (h-BN) counterparts (where similar thermal annealing does not cause as much measurable enhancement),[40,41] likely partially due to the relatively stronger affinity of β-Ga$_2$O$_3$ with gaseous adsorbates in ambient environment. The post-annealing measured upshifts in $f$s and enhanced $Q$s can be qualitatively explained by the annealing enabled cleaning of possible adsorbates and residues (lowering the mass $M$ in $\omega = \sqrt{k/M}$), and the alleviation or elimination of their associated dissipation and energy loss processes. Most of the measured resonances show decreased dissipation rates ($f/Q$) after annealing (see Table S2), further confirming the alleviation of dissipation pathways by thermal annealing.

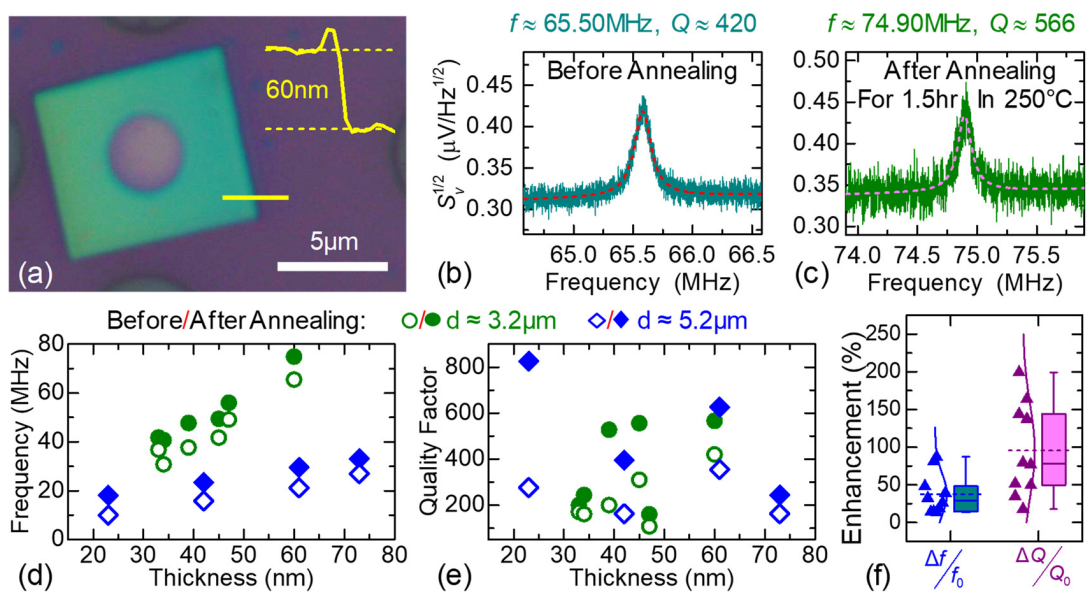

**Figure 2.** Measurement results of β-Ga$_2$O$_3$ diaphragm nanoresonators before and after thermal annealing. (a) Optical microscopy image of a typical β-Ga$_2$O$_3$ diaphragm nanoresonator with a diameter of $d \approx 3.2$ μm and a thickness of $t \approx 60$ nm. *Inset*: Atomic force microscopy (AFM) trace of the corresponding cross section labeled in the optical image. (b) & (c) Thermomechanical noise spectra of the device's fundamental mode, before and after a 250 °C, 1.5 hour thermal annealing, respectively. (d) Fundamental mode resonance frequencies ($f_0$) and (e) quality ($Q$) factors of the β-Ga$_2$O$_3$ diaphragm nanoresonators before and after annealing. Hollow symbols show results before annealing and solid ones show the results after annealing. Blue diamonds represent devices with diameter $d \approx 5.2$ μm and green circles represent $d \approx 3.2$ μm devices. (f) Measured post-annealing enhancement (in percentage) and dispersion of $f_0$ and $Q$.

To quantitatively understand the measured resonances, we first perform analysis on multimode resonances and frequency scaling for these circular β-Ga$_2$O$_3$ diaphragms. In any given circular drumhead device, both flexural rigidity (dominated by thickness and elastic modulus) and built-in tension (stress) can be important, thus we have,[42]

$$\omega_m = (k_m a) \sqrt{\frac{D}{\rho \cdot t \cdot a^4} \left[ (k_m a)^2 + \frac{\gamma \cdot a^2}{D} \right]},$$  (1)

where $m$ denotes the mode number, $\omega_m = 2\pi f_m$ is the $m^{\text{th}}$ mode angular resonance frequency, $(k_m a)^2$ is the eigenvalue which can be numerically calculated ($k_0 a = 10.215$, $k_1 a = 21.260$, $k_2 a = 34.877$, $k_3 a = 39.771$),[42] $a$ is the radius of circular resonator, $D$ is the flexural rigidity $D = E_{\text{Y}} t^3 / \left[ 12(1 - \nu^2) \right]$ with $t$ being the thickness of the device, $\nu$ is the Poisson's ratio, $\rho$ is the volume mass density of β-Ga$_2$O$_3$, and $\gamma$ is the surface pre-tension evenly distributed in the plane. Equation (1) yields a 'mixed elasticity' model that captures both the 'disk' and 'membrane' limits of frequency scaling for such devices, as well as the transition between these two regimes.[43] In the 'disk' regime where resonance frequency is dominated by the flexural rigidity, large $D$ makes the $\gamma a^2/D$ term in Equation (1) negligible. The resonance frequencies are positively dependent on resonator thickness, $\omega_m \propto t$. In the 'membrane' regime where the built-in tension dominates, the term $\gamma a^2/D \gg (k_m a)^2$, and $(k_m a)^2$ term can be neglected. The resonance frequencies scale with resonator thickness as $\omega_m \propto t^{1/2}$. For circular drumhead resonators in the 'disk' regime, the built-in tension has negligible effect on determining the resonance frequency. Therefore, if the device is in the 'disk' regime, the Young's modulus can be revealed from measured fundamental-mode resonance frequency $\omega_0$ by

$$E_{\text{Y}} = \frac{12 a^4 \rho \left( 1 - \nu^2 \right)}{\left[ \left( k_0 a \right)^2 \right]^2 \cdot t^2} \omega_0^2,$$  (2)

with $\left( k_0 a \right)^2 = 10.215$ for the fundamental mode. We choose the applicable devices for Young's modulus extraction. By fitting and plotting the measured data to the $\omega_0 \propto t$ relation (straight dimmed solid lines in Figure 3a), we can compare the measured resonance data with the ideal 'disk' curve. We exclude the 2 devices deviated from the ideal 'disk' curve (2 thinner devices with diameter of ~5.2 μm) along with devices with non-ideal geometry (see Figure S5 in SI) for extracting $E_Y$. Using Equation (2) and the measured fundamental mode resonances of the rest devices, we extract an average Young's modulus, $E_Y \approx 261$ GPa, for the β-Ga$_2$O$_3$ nanoflakes.

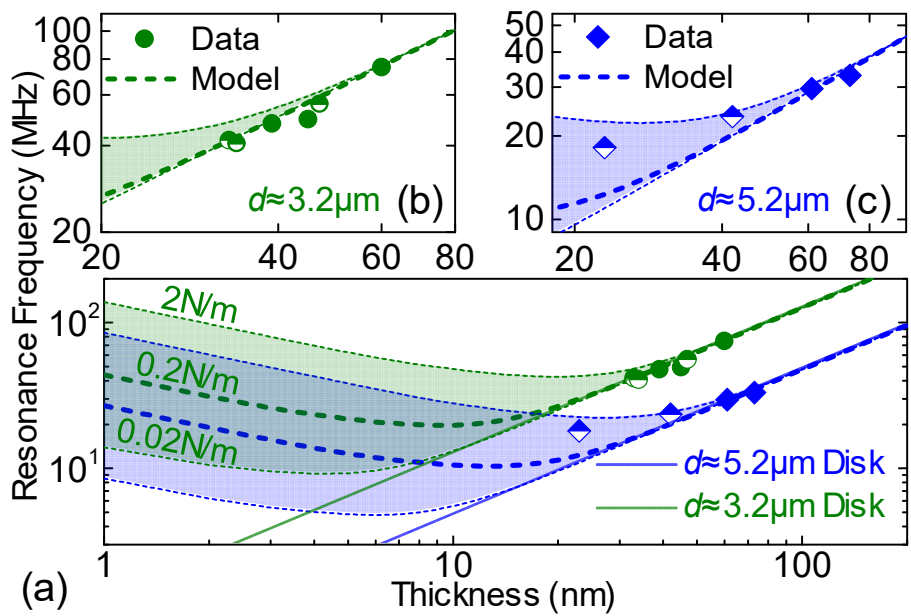

**Figure 3**. Modeling and elucidating the mixed elastic behavior and transition from the 'disk' regime to the 'membrane' regime in β-Ga$_2$O$_3$ diaphragm nanoresonators. Green circles and blue diamonds represent devices with diameters of $d \approx 3.2$ μm and $d \approx 5.2$ μm, respectively. Scattered symbols represent experimental data from optical interferometry measurements with solid ones representing devices used for Young's modulus extraction. Curved lines show calculated resonance frequency *vs* device thickness, each with 0.02 N m$^{-1}$, 0.2 N m$^{-1}$ and 2 N m$^{-1}$ surface tensions, plotted using Equation (1). Dimmed solid lines indicate the relation for ideal 'disk' resonators. (a) Frequency scaling of β-Ga$_2$O$_3$ diaphragm nanoresonators with thicknesses in 1nm to 100 nm range. (b) & (c) Zoomed-in scaling of measured devices' thickness ranges for resonators with diameters of $d \approx 3.2$ μm and $d \approx 5.2$ μm, respectively.

By applying the measured Young's modulus to the resonator model, we complete the frequency scaling of β-Ga$_2$O$_3$ nanomechanical resonators from the 'membrane' regime to the 'disk' regime (Figure 3). The plots confirm that all the resonators with ~3.2 μm diameter lie in the 'disk' regime. For the resonators with ~5.2 μm diameter, the thicker 2 devices lie in the 'disk' regime while the other 2 resonators lie in the transition regime.

To gain comprehensive understanding of the mechanical properties and resonance performance of the β-Ga$_2$O$_3$ diaphragms, we further investigate the multimode resonances and spatial mapping of their mode shapes. Figure 4 displays the measured results from a resonator with a diameter $d \approx$ 5.7 μm and thickness $t \approx$ 23 nm, which, according to frequency scaling, lies in the transition regime. We measure up to the 6th undriven thermomechanical resonance mode, including resonance spectra and mapping of mode shapes, both before and after annealing. By vividly discerning and visualizing the multimode shapes, we can easily compare and match the resonance modes before and after annealing. Similar to the fundamental-mode resonances, the higher order modes also exhibit clear upshifts in $f$s and enhancement of $Q$s after annealing. The rich resonance modes indicate that this device can potentially be used as the platform for studying the parametric pumping and modes coupling of β-Ga2O3 nanomechanical resonators.[44,45]

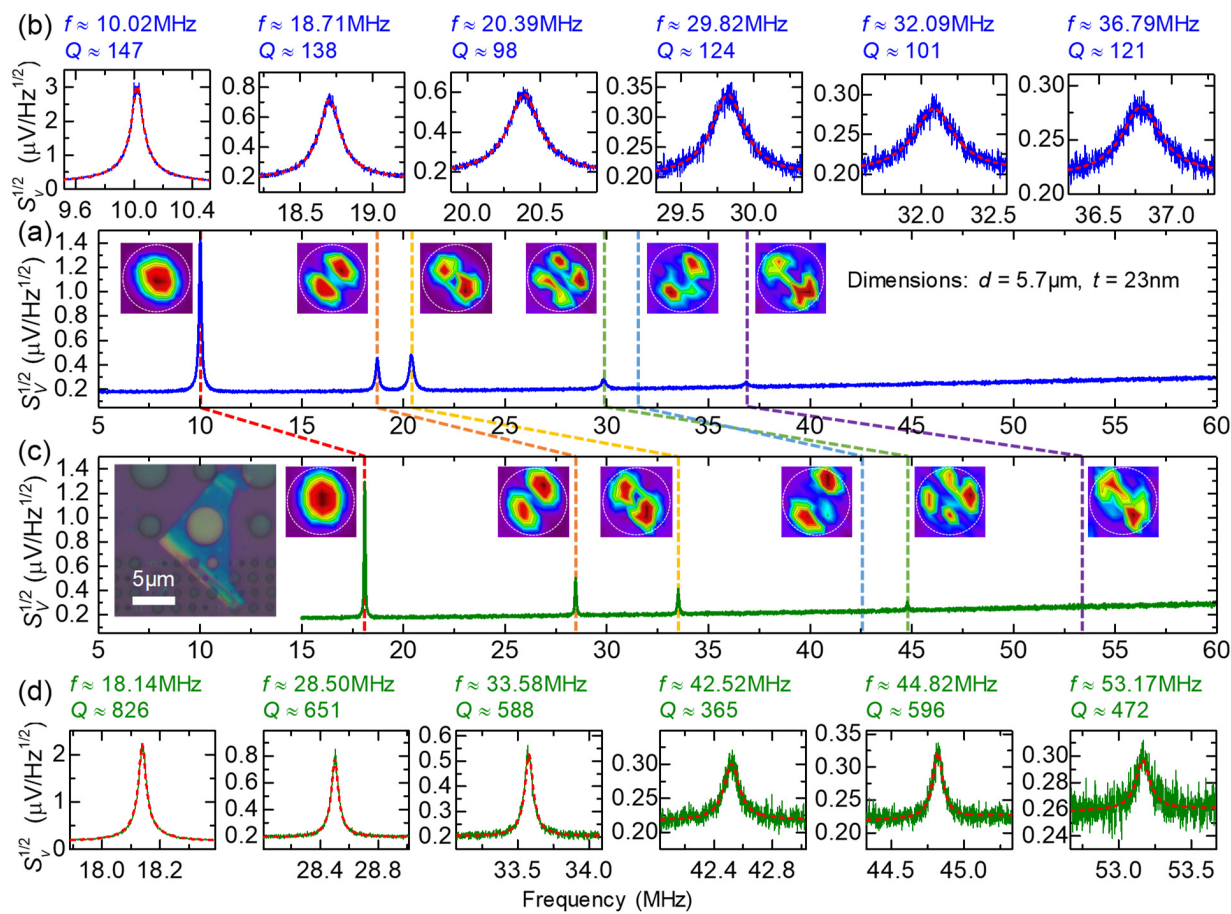

**Figure 4.** Comparison of measured multimode resonances before and after thermal annealing of a β-Ga2O3 diaphragm nanoresonator (device optical image shown in the *inset* of panel (c)). Thermomechanical resonance spectra of (a) before and (c) after thermal annealing (250 °C, 1.5 hrs) with zoomed-in individual resonance spectra with fitted $f$s and $Q$s (panels in (b) and (d), respectively), and spatial mapping and visualization of the resonance mode shapes (*insets* of (a) and (c)).

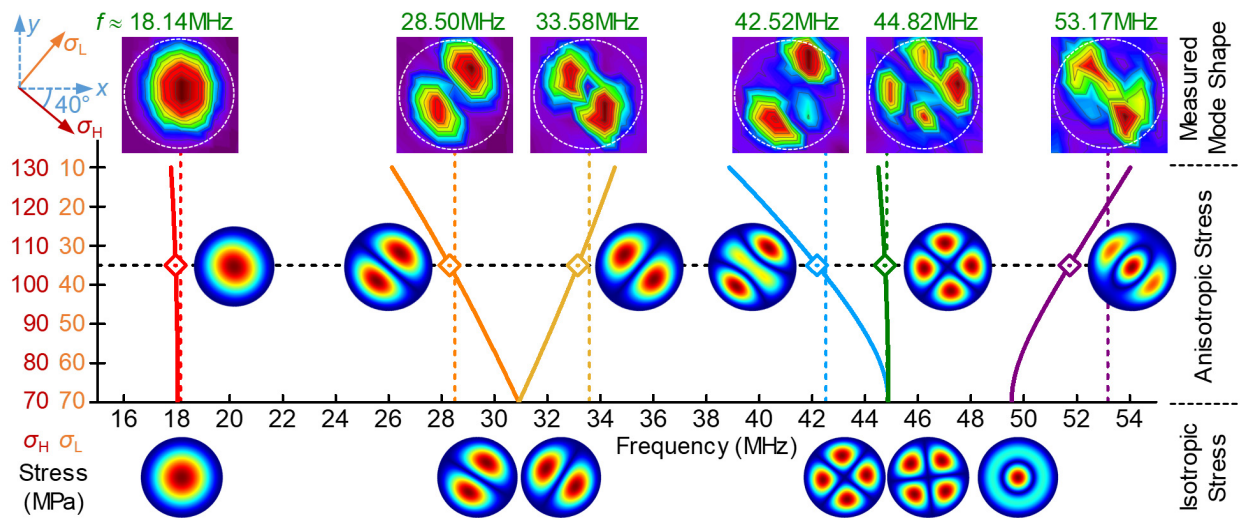

**Figure 5.** Determination of anisotropic tension in the circular β-Ga$_2$O$_3$ diaphragm nanoresonator in Figure 4 by matching the simulated mode shapes and frequencies with the spatially resolved and visualized multimode resonances. The simulation is started using the resonator model with isotropic stress $\sigma$ = 70 MPa ($\gamma$ = 1.61 N m$^{-1}$, last row of the figure). By separating the stress in different directions in simulation, the results with biaxial stress of $\sigma_H$ = 105 MPa ($\gamma_H$ = 2.42 N m$^{-1}$) and $\sigma_L$ = 35 MPa ($\gamma_L$ = 0.81 N m$^{-1}$) best match the measured results, where $\sigma_H$ direction is 40° off from the horizontal direction.

Importantly, the multimode spatial mapping reveals mechanical properties and features that could be otherwise not visible or measurable, thus shedding more light on understanding these circular drumhead resonators. As shown in Figure 3c, the multimode device measured in Figure 4 ($d \approx 5.7$ μm, $t \approx 23$ nm) behaves in the transition regime, where both built-in tension and flexural rigidity are important in determining its resonance frequencies. When we apply a uniform built-in tension and sweep this tension value in FEM simulations, it yields a uniform stress of $\sigma$ = 70 MPa (1.61 N m$^{-1}$ surface tension) that matches the fundamental-mode resonance frequency, whereas the simulated frequencies of higher order modes include well-known degenerate mode pairs that have theoretically identical, non-distinguishable frequencies (*i.e.*, see modes 2 & 3, modes 4 & 5 in the bottom row of Figure 5). In clear contrast, measured resonances labeled with vertical dashed lines in Figure 5 show salient splitting in the degenerate mode pairs, indicating clear broken symmetry. This suggests that the built-in tension cannot be uniform, but anisotropic instead. Such an in-plane anisotropy is also strongly suggested by the spatially mapped multimode shapes demonstrated in the top row of Figure 5, where the nodal lines in the mapped shapes of mode 2 and 3 indicate the stiffer and softer axis of in-plane anisotropy, respectively. To be quantitative, we sweep the anisotropic, biaxial built-in tension in modeling. We find that when the biaxial built-in tensions are 105 MPa and 35 MPa (corresponding to 2.42 N m$^{-1}$ and 0.81 N m$^{-1}$ of surface tension, respectively), the simulated resonances are in good agreement with the measured multimode responses, for both the frequencies and the mode shapes. We further increase the stress in both directions in simulation to minimize the discrepancy between the simulated resonance frequencies and measured ones. A set of slightly increased biaxial stress, $\sigma_H = 107.5$ MPa ($\gamma_H = 2.47$ N m$^{-1}$) and $\sigma_L = 37.5$ MPa ($\gamma_L = 0.86$ N m$^{-1}$), generates simulation results ($f_1 = 18.18$ MHz, $f_2 = 28.68$ MHz, $f_3 = 33.47$ MHz, $f_4 = 42.77$ MHz, $f_5 = 45.30$ MHz, $f_6 = 52.28$ MHz) that match excellently with the measured resonance frequencies ($f_1 = 18.14$ MHz, $f_2 = 28.50$ MHz, $f_3 = 33.58$ MHz, $f_4 = 42.52$ MHz, $f_5 = 44.82$ MHz, $f_6 = 53.17$ MHz).

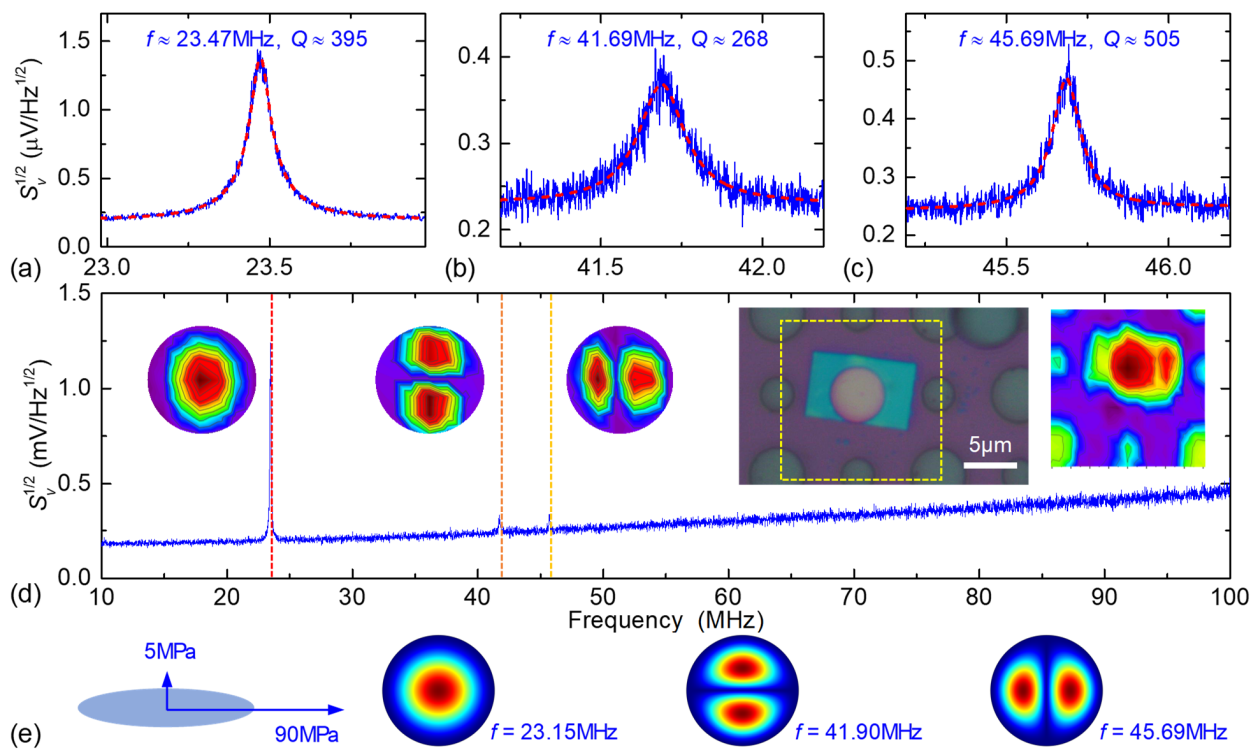

**Figure 6.** Resonances of a 42nm thick β-Ga$_2$O$_3$ nanomechanical resonator. Zoomed-in measured resonance spectra of (a) 1$^{st}$, (b) 2$^{nd}$ and (c) 3$^{rd}$ resonance mode, and resonance frequencies and quality factors are labeled. (d) Wide-range (10MHz-100MHz) spectrum of this device with the thermomechanical motion maps of each resonance mode shape visualized by laser interferometry scanning spectromicroscopy. *Insets*: optical microscopy image of the device with the corresponding static interferometry map. (e) Simulated mode shapes for the device with anisotropic built-in stress.

In another thicker, transition regime device ($t \approx 42$ nm, $d \approx 5.2$ μm) with 3 measured resonance modes, multimode resonances with spatial mapping exhibit similar mode splitting features that reveal in-plane anisotropy in elastic behavior (Figure 6). Once again, parametric sweeping modeling yields an anisotropic built-in tension (5 MPa and 90 MPa, corresponding to 0.21 N m$^{-1}$ and 3.78 N m$^{-1}$ of surface tension) that excellently reproduces the measured mode splitting features. These newly resolved non-uniform, anisotropic components of built-in tension in these devices are very reasonable and intuitively understandable, as they can be likely introduced during the nanoflake dry-transfer process where the axis – along which the polymeric tape is pressed down to make contact to substrate and is then lifted up – is likely to build a higher stress component than along its perpendicular direction.

In conclusion, we have demonstrated a new type of UWBG nanomechanical resonator based on synthesized single-crystal β-Ga$_2$O$_3$ nanoflakes. These β-Ga$_2$O$_3$ nanoresonators demonstrate robust multimode resonances in the high frequency (HF) and very high frequency (VHF) bands ranging from 18 MHz to 75 MHz. From devices operating in the 'disk' regime, the measured resonances determine a Young's modulus of $E_Y \approx 261$ GPa. Multimode spatial mapping discloses new mode splitting features, which in combination with parametric modeling, reveal anisotropic built-in tension (at the levels of 0.2−3.8 N m$^{-1}$) in devices that operate in the transition regime. Not only does this show nanomechanical resonances with multimode spatial mapping are a powerful probe for revealing device properties that are otherwise invisible or unmeasurable, but also it helps pave the way for future nanomechanical engineering of β-Ga$_2$O$_3$ crystals toward electromechanical and opto-electromechanical devices that harness the crystal's unique properties, which will greatly supplement the rapidly emerging β-Ga$_2$O$_3$ electronic devices.

## Methods

### β-Ga₂O₃ Nanoflake LPCVD Synthesis

The construction of β-Ga$_2$O$_3$ drumhead resonators starts with the synthesis of low-dimensional β-Ga$_2$O$_3$ nanoflakes by using a low-pressure chemical vapor deposition (LPCVD) method.[34]  The nanoflakes are synthesized on a 3C-SiC epi-layer on a Si substrate in a conventional tube furnace with programmable gas and temperature controllers.  Over a 1.5 hour period at a growth temperature of 950 °C, using high purity Ga pellets (Alfa Aesar, 99.99999%) as source material and O$_2$ as a gaseous precursor, the formation of β-Ga$_2$O$_3$ proceeds, step by step, from β-Ga$_2$O$_3$ nanocrystals to β-Ga$_2$O$_3$ nanorods, and then nanoflakes, without any foreign catalyst (Figure S1 in SI).  The as-grown nanoflakes have widths of ~2−30 μm and thicknesses of ~20−140 nm.

### Suspended Diaphragm Fabrication

Large arrays of microtrenches are lithographically defined and fabricated on a 290 nm-SiO$_2$-on-Si substrate.  The SiO$_2$ layer is patterned with circular microtrenches using photolithography, and then etched by reactive ion etching (RIE).  The depth of resulting SiO$_2$ microtrenches is 290 nm in this experiment, and the designed nominal values for the microtrench diameters are 3 μm and 5 μm (corresponding to actual diameters of ~3.2 μm and ~5.2 μm measured after the fabrication process).

   Using the synthesized β-Ga$_2$O$_3$ nanoflakes, we fabricate suspended β-Ga$_2$O$_3$ nanostructures by employing an all-dry transfer technique (Figure S2 in SI).  With the assistance of thermal release tape, we pick up the β-Ga$_2$O$_3$ nanoflakes from the as-grown samples and thermally release the flakes onto the pre-fabricated 290 nm-SiO$_2$-on-Si substrate with pre-defined circular microtrenches.  Using this method, we fabricate β-Ga$_2$O$_3$ circular drumhead resonators with thicknesses of ~20 to ~80 nm and diameters of ~3.2 μm and ~5.2 μm.

### Atomic Force Microscopy (AFM) and Raman Spectroscopy

AFM measurements are conducted with an Agilent 5500 AFM using tapping mode.  Raman measurements are performed using a customized micro-Raman system that is integrated into the scanning laser interferometric resonance measurement system (Figure 1d).  A 532 nm laser is focused on the transferred β-Ga$_2$O$_3$ flake using a 50× microscope objective. The typical laser spot size is ∼1 μm. Raman scattered light from the sample is collected in backscattering geometry and then guided to a spectrometer (Horiba iHR550) with a 2400 g mm$^{-1}$ grating and 2 min integration time. The Raman signal is recorded using a liquid-nitrogen-cooled CCD. The spectral resolution of our system is ∼1 cm$^{-1}$.

**Scanning Laser Interferometry and Spectromicroscopy**

Resonances of the circular β-Ga$_2$O$_3$ drumhead resonators are characterized using scanning laser interferometry detection techniques (Figure 1d). A 633 nm He-Ne laser is focused on the device surface for detection of the motion-modulated interference between multiple reflections from the vacuum-flake, flake-cavity and cavity-Si interfaces, which is read out by a low-noise photodetector. A spectrum analyzer is used to measure the signal from the photodetector. We apply an incident laser power of ∼2.8 mW to the devices which are preserved under moderate vacuum conditions (∼20 mTorr) to ensure reasonable signal-to-noise ratio while avoiding overheating. For a given resonance mode, performing the scanning spectromicroscopy measurements yields the spatial mapping and visualization of the mode shape of this resonance. All the interferometry measurements in this work are done at room temperature.

**Displacement Sensitivity**

We calculate the resonator thermomechanical noise in the displacement domain using

$$S_{x,\text{th}}^{1/2}(\omega) = \left( \frac{4k_{\text{B}}T\omega_m}{M_{m,\text{eff}} \cdot Q_m} \cdot \frac{1}{\left( \omega^2 - \omega_m^2 \right)^2 + \left( \omega\omega_m/Q_m \right)^2} \right)^{1/2}. \tag{3}$$

When the device is on resonance ($\omega = \omega_m$), the expression simplifies to

$$S_{x,\text{th}}^{1/2}(\omega_m) = \left( \frac{4k_{\text{B}}TQ_m}{M_{m,\text{eff}} \cdot \omega_m^3} \right)^{1/2}, \tag{4}$$

where $\omega_m$, $k_B$, $T$, $Q_m$, $M_{m,\text{eff}}$ are angular resonance frequency, Boltzmann's constant, temperature, quality factor, and effective mass, respectively, and $m$ denotes the $m^{\text{th}}$ mode of the resonator. Effective mass can be calculated using

$$M_{m,\text{eff}} = \frac{\iint_S z_m^2(x,y)\,dS}{\pi a^2 \cdot z_{m,\max}^2} M \,, \tag{5}$$

where $z_m(x,y)$ is the mode shape of the $m^{\text{th}}$ mode, $z_{m,\max}$ is the maximum displacement, and $M$ is the mass of the resonator. We assume that laser heating is minimal due to the UWBG nature of β-Ga$_2$O$_3$ and therefore the temperature of the sample remains at room temperature ($T \approx 300$ K). The effective mass of the resonator in the 'disk' regime is $M_{\text{eff}} = 0.1828M$. The effective mass of the transition regime devices are calculated by incorporating FEM simulated mode shapes into Equation (5).

Using this analysis, the displacement spectral density of the resonator can be calculated. Hence, we determine the responsivity of the laser interferometry system $\Re \equiv S_{v,\text{th}}^{1/2} / S_{x,\text{th}}^{1/2}$ and system sensitivity $S_{x,\text{sys}}^{1/2} \equiv S_{v,\text{sys}}^{1/2} / \Re$ for each device. Here, $S_{v,\text{th}}^{1/2}$ and $S_{v,\text{sys}}^{1/2}$ are the voltage domain thermomechanical noise and the noise level of measurement system, respectively. For all devices in this work, the responsivities and sensitivities of the interferometry system range from 12 μV pm$^{-1}$ to 60 μV pm$^{-1}$ and from 4 fm Hz$^{-1/2}$ to 19 fm Hz$^{-1/2}$, respectively.

**Supporting Information**: Material synthesis and resonator fabrication processes; Raman spectroscopy for crystal quality characterization; optical interferometry displacement sensitivity analysis; Young's modulus extraction and resonator frequency scaling; summary of devices and their measured parameters.

**Acknowledgement**: We thank financial support from the U.S. Department of Energy (DOE) EERE Grant (DE-EE0006719) and the Great Lakes Energy Institute (GLEI) ThinkEnergy Fellowship. C. A. Zorman and P. X-L. Feng thank financial support from the National Science Foundation (NSF) SNM Program (Grant CMMI-1246715). S. Rafique and H. Zhao thank financial support from the NSF DMR Grant (1755479). L. Han and H. Zhao thank financial support from the NSF CNS Grant (1664368). We thank J. P. McCandless for helpful discussions. Part of the device fabrication was performed at the Cornell NanoScale Science and Technology Facility (CNF), a member of the National Nanotechnology Infrastructure Network (NNIN), supported by the National Science Foundation (Grant ECCS-0335765).

*– Supporting Information –*

# Ultrawide Band Gap β-Ga₂O₃ Nanomechanical Resonators

# with Spatially Visualized Multimode Motion

Xu-Qian Zheng[1], Jaesung Lee[1], Subrina Rafique[1], Lu Han[1],

Christian A. Zorman[1], Hongping Zhao[1], Philip X.-L. Feng[1,*]

[1]*Department of Electrical Engineering & Computer Science, Case School of Engineering,*

*Case Western Reserve University, 10900 Euclid Avenue, Cleveland, OH 44106, USA*

## Table of Contents

**Materials and Methods**

*Corresponding Author. Email: philip.feng@case.edu

## S1. Material Synthesis and Resonator Fabrication

The construction of β-Ga₂O₃ drumhead resonators starts with the synthesis of single-crystal β-Ga₂O₃ nanoflakes by using a low-pressure chemical vapor deposition (LPCVD) method.[1]  The nanoflakes are synthesized on top of a 3C-SiC epi-layer on Si substrate in a tube furnace with programmable gas and temperature controllers.  We use high purity Ga pellets (Alfa Aesar, 99.99999 %) as source material and O₂ as a gaseous precursor.  With a growth temperature of 950 °C for 1.5 hours, the formation of β-Ga₂O₃ proceeds, step by step, from β-Ga₂O₃ nanocrystals to β-Ga₂O₃ nanorods, and then nanoflakes, without any foreign catalyst (Figure S1).  The as-grown nanoflakes have widths of ~2−30 μm and thicknesses of ~20−140 nm.

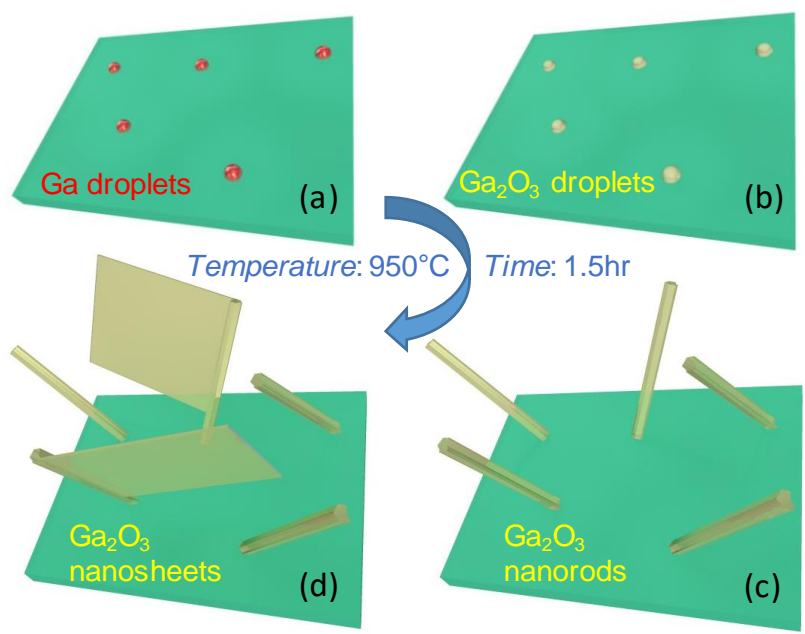

**Figure S1**.  Schematic illustration of the LPCVD β-Ga₂O₃ nanoflake synthesis process.  While maintaining 950 °C temperature in evaporated Ga and gaseous O₂/Ar environment for 1.5 hours, the β-Ga₂O₃ nanoflakes form on 3C-SiC-on-Si substrate, in the sequence of (a) adsorption of Ga adatoms and gathering of Ga droplets, (b) formation of β-Ga₂O₃ droplets through oxidation, (c) extrusion of β-Ga₂O₃ nanorods due to supersaturation, and (d) nucleation and extrusion of β-Ga₂O₃ nanoflakes from nanorod sidewalls.

Using the synthesized β-Ga₂O₃ nanoflakes, we fabricate suspended β-Ga₂O₃ nanostructures by employing an all-dry transfer technique (Figure S2). With the assistance of thermal release tape, we pick up the β-Ga₂O₃ nanoflakes from the as-grown samples and thermally release the flakes to a pre-fabricated 290 nm-SiO₂-on-Si substrate with pre-defined arrays of circular microtrenches. Using this method, we fabricate β-Ga₂O₃ circular drumhead resonators with thicknesses of ~20 to ~80 nm and diameters of ~3.2 μm and ~5.2 μm.

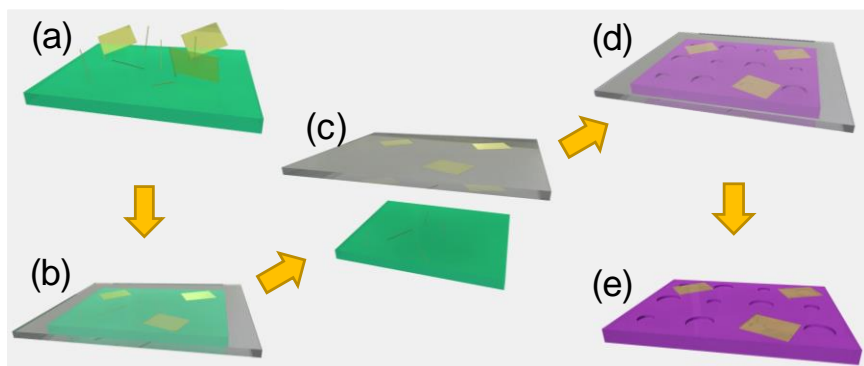

**Figure S2**. Illustration of the all-dry transfer process for fabrication of circular drumhead β-Ga₂O₃ nanomechanical resonators. (a) Synthesized β-Ga₂O₃ nanorods and nanoflakes on 3C-SiC epi-layer. (b) Application of thermal release tape. (c) Picking up β-Ga₂O₃ nanoflakes by thermal release tape. (d) Applying the tape with flakes to desired locations on the substrate and releasing nanoflakes at >90 °C temperature. (e) Fabricated β-Ga₂O₃ drumhead nanoresonators.

## S2. Raman Spectroscopy

To characterize the crystalline properties of the β-Ga$_2$O$_3$ nanoflakes after transfer, we examine Raman scattering signatures of suspended β-Ga$_2$O$_3$ flakes using a 532nm laser. We find Raman modes at 143 cm$^{-1}$, 168 cm$^{-1}$, 199 cm$^{-1}$, 346 cm$^{-1}$, 416 cm$^{-1}$, 476 cm$^{-1}$, 653 cm$^{-1}$, and 768 cm$^{-1}$ (Figure S3) from our devices, which very well match Raman modes of bulk β-Ga$_2$O$_3$,[2] showing high crystal quality of the β-Ga$_2$O$_3$ nanoflakes after the transfer process.

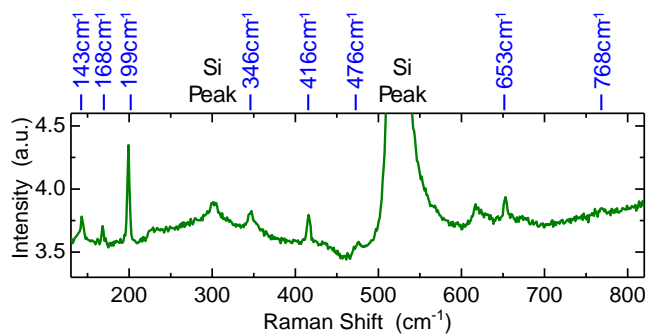

**Figure S3**. A typical Raman spectrum measured from a suspended β-Ga$_2$O$_3$ drumhead device.

To further study the effects of thermal annealing on the crystalline properties of β-Ga$_2$O$_3$ resonators, we measure the Raman spectrum of a β-Ga$_2$O$_3$ flake at two different locations both before and after annealing (Figure S4). The thermal annealing is performed at 250 °C in a vacuum of ~500 mTorr for 1.5 hours. To precisely determine the Raman signatures, we fit each peak with a Lorentzian curve. Table S1 summarizes the Raman peak positions and values of full width at half maximum (FWHM) for four β-Ga$_2$O$_3$ Raman peaks measured from the fitted spectra in Figure S4. We find that the Raman peak positions after annealing are consistent with those before annealing: peak shifts are within the spectral resolution of our measurement system (~1 cm$^{-1}$), indicating the change of strain level, if any, is very small and below the strain sensitivity of Raman spectroscopy.

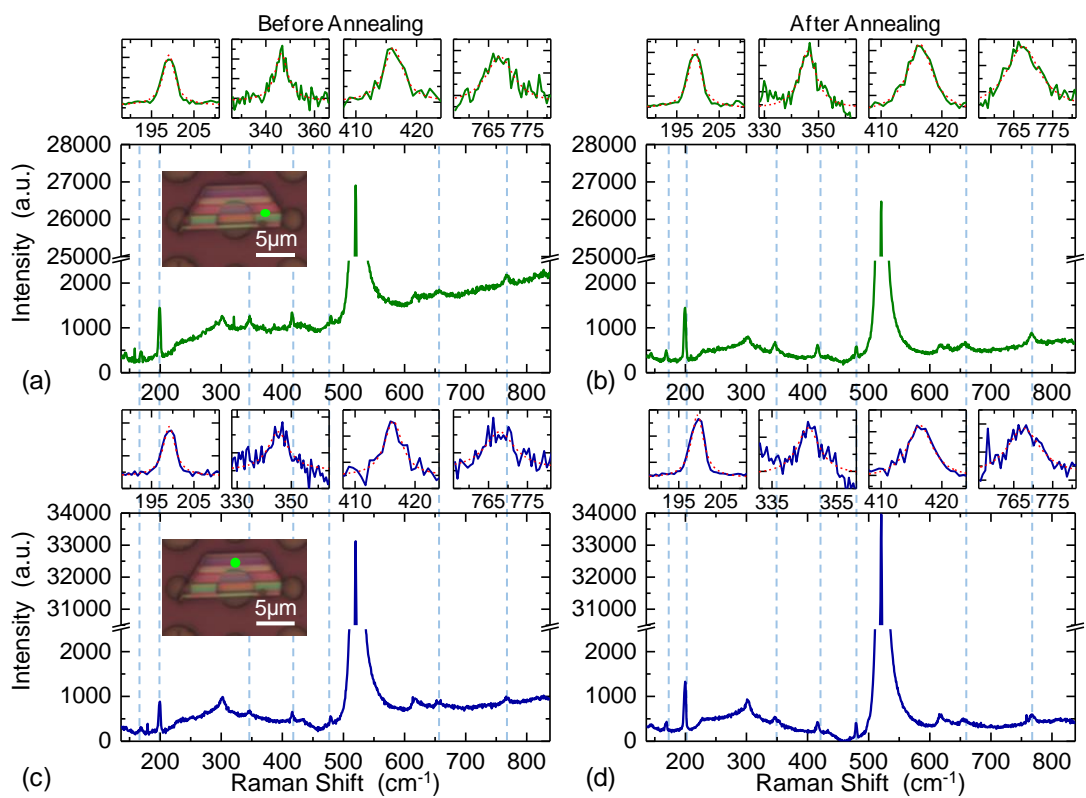

**Figure S4.** Raman spectra measured from a dry-transferred β-Ga₂O₃ nanoflake (a)&(c) before and (b)&(d) after thermal annealing, with zoomed-in spectra showing fittings for full width at half maximum (FWHM) of Raman modes $A_g^{(3)}$ (199 cm⁻¹), $A_g^{(5)}$ (346 cm⁻¹), $A_g^{(6)}$ (416 cm⁻¹) and $A_g^{(10)}$ (768 cm⁻¹). *Insets* show the optical images of the device with the measurement spots of the Raman spectroscopy.

**Table S1**. Measured Raman Characteristics before and after Thermal Annealing

| Laser Spot | Raman Mode | Before Annealing | | After Annealing | |
|---|---|---|---|---|---|
| | | Peak Position (cm⁻¹) | FWHM (cm⁻¹) | Peak Position (cm⁻¹) | FWHM (cm⁻¹) |
| #1 | $A_g^{(3)}$ | 199.11 | 3.44 | 199.26 | 3.56 |
| | $A_g^{(5)}$ | 346.32 | 6.30 | 346.04 | 7.48 |
| | $A_g^{(6)}$ | 416.20 | 3.80 | 416.39 | 5.02 |
| | $A_g^{(10)}$ | 767.90 | 8.72 | 767.38 | 10.49 |
| #2 | $A_g^{(3)}$ | 199.18 | 3.39 | 199.42 | 3.53 |
| | $A_g^{(5)}$ | 345.14 | 8.99 | 346.16 | 8.50 |
| | $A_g^{(6)}$ | 416.38 | 3.62 | 416.40 | 5.05 |
| | $A_g^{(10)}$ | 767.86 | 9.36 | 767.54 | 10.33 |

## S3. Displacement Sensitivity

We analyze the measured Brownian motion of devices to calibrate the responsivity of the laser interferometry system from mechanical displacement to electrical signal. We calculate the thermomechanical noise in displacement domain using

$$S_{x,\text{th}}^{1/2}(\omega) = \left( \frac{4k_B T \omega_m}{M_{m,\text{eff}} \cdot Q_m} \cdot \frac{1}{\left(\omega^2 - \omega_m^2\right)^2 + \left(\omega \omega_m / Q_m\right)^2} \right)^{1/2}. \tag{S1}$$

When the device is on resonance ($\omega = \omega_n$), the expression is simplified to

$$S_{x,\text{th}}^{1/2}(\omega_m) = \left( \frac{4k_B T Q_m}{\omega_m^3 M_{m,\text{eff}}} \right)^{1/2}, \tag{S2}$$

where $\omega_m$, $k_B$, $T$, $Q_m$, $M_{m,\text{eff}}$ are angular resonance frequency, Boltzmann's constant, temperature, quality factor, and effective mass, respectively, and $m$ denotes the $m^{\text{th}}$ mode of the resonator.

To derive the effective mass for a circular diaphragm resonator with clamped edge, we start from calculating the kinetic energy of the system:

$$E = \iint_m \frac{1}{2} u_m^2 dm, \tag{S3}$$

where $u_m$ is the velocity distribution of the $m^{\text{th}}$ mode of the resonator, which can be given by

$$u_m(x, y) = \frac{z_m(x, y)}{z_{m,\text{max}}} \dot{z}, \tag{S4}$$

where $z_m(x,y)$ is the mode shape of the $m^{\text{th}}$ mode, $z_{m,\text{max}}$ is the maximum displacement of the $m^{\text{th}}$ mode, and $\dot{z}$ is the velocity of the device in a lumped-parameter fashion. By substituting Equation (S4) into Equation (S3), we have

$$E = \frac{1}{2} M \cdot \frac{\iint_S z_m^2(x, y) dS}{\pi a^2 \cdot z_{m,\text{max}}^2} \cdot \dot{z}^2 = \frac{1}{2} M_{m,\text{eff}} \dot{z}^2, \tag{S5}$$

where $M$ is the device mass, $a$ is the radius, and $M_{m,\text{eff}}$ is the effective mass of the $m^{\text{th}}$ mode of the resonator. From Equation (S5), effective mass can be calculated by

$$M_{m,\text{eff}} = \frac{\iint_S z_m^2(x,y)\,dS}{\pi a^2 \cdot z_{m,\text{max}}^2} M \ . \tag{S6}$$

For devices the in 'disk' regime, the deflection for the fundamental mode as a function of the radial position $r$ $(0 \leq r \leq a)$ is given by[3]

$$z_0(r) = J_0(k_0 r) - \frac{J_0(k_0 a)}{I_0(k_0 a)} \cdot I_0(k_0 r), \tag{S7}$$

where $J_0$ is the $0^{\text{th}}$-order Bessel function $J$, $I_0$ is the $0^{\text{th}}$-order extended Bessel function $I$, and $(k_m a)^2$ is the eigenvalue (*e.g.*, $(k_0 a)^2 = 10.215$ for the fundamental mode of the resonator in the 'disk' limit). By plugging Equation (S7) into Equation (S6), the fundamental mode effective mass of a resonator in the 'disk' regime is $M_{0,\text{eff}} = 0.1828M$. For the devices falling in the transition regime, we calculate the effective masses by incorporating the mode shapes, computed by using finite element method (FEM, in COMSOL Multiphysics) simulations, into Equation (S6).

We assume laser heating is minimal because the photon energy of 633nm laser (1.96eV) is much lower than the bandgap of β-Ga₂O₃ (4.5-4.9eV) and, therefore, device temperature remains at room temperature ($T \approx 300$K). Using aforementioned analysis from Equation (S1) to (S7), displacement domain thermomechanical noise $S_{x,\text{th}}^{1/2}$ of the resonator can be calculated. Further, we determine the responsivity $\Re \equiv S_{v,\text{th}}^{1/2} / S_{x,\text{th}}^{1/2}$ and displacement sensitivity of the laser interferometry system $S_{x,\text{sys}}^{1/2} \equiv S_{v,\text{sys}}^{1/2} / \Re$ for each device. Here, $S_{v,\text{th}}^{1/2}$ and $S_{v,\text{sys}}^{1/2}$ are voltage domain thermomechanical noise and noise level of the measurement system, respectively. For all devices in this work, the responsivities and sensitivities of the interferometry system range from 12 μV/pm to 60 μV/pm and from 4 fm/Hz$^{1/2}$ to 19 fm/Hz$^{1/2}$, respectively.

## S4. Young's Modulus Extraction and Resonance Frequency Scaling

We analyze the resonance frequency scaling of the β-Ga$_2$O$_3$ circular drumhead resonators. The resonance frequency of a circular drumhead resonator is

$$\omega_m = \left(k_m a\right)\sqrt{\frac{D}{\rho \cdot t \cdot a^4}\left[\left(k_m a\right)^2 + \frac{\gamma \cdot a^2}{D}\right]}, \tag{S8}$$

where $a$ is the radius of circular resonator, $D$ is the flexural rigidity $D = E_\gamma t^3 / \left[12\left(1-\nu^2\right)\right]$, $\rho$ is the volume mass density of β-Ga$_2$O$_3$, and $\gamma$ is the surface pre-tension evenly distributed in the plane.[4] The eigenvalue $(k_m a)^2$ can be calculated using the equation

$$\left(k_m a\right)^2 = \alpha + \left(\beta - \alpha\right)e^{-\eta \exp\{\delta \ln(x)\}}, \tag{S9}$$

where $x = \gamma a^2 / D$ and for different modes, the eigenvalues are calculated using numerically calculated parameters $\alpha$, $\beta$, $\eta$ and $\delta$ given in Ref. 4. In addition, as if the device is ideal 'membrane' or ideal 'disk', this eigenvalue equals $\alpha$ or $\beta$, respectively.

In the limit that flexural rigidity dominates, the large flexural rigidity $D$ makes $\gamma a^2/D$ term in Equation (S8) diminish and Equation (S8) becomes

$$\omega_m = \frac{\left(k_m a\right)^2}{a^2}\sqrt{\frac{D}{\rho \cdot t}}. \tag{S10}$$

Then, Young's modulus can be determined by

$$E_Y = \frac{12a^4 \rho \left(1-\nu^2\right)}{\left[\left(k_0 a\right)^2\right]^2 \cdot t^2}\omega_0^2, \tag{S11}$$

from the fundamental mode resonance frequency. Here, $(k_0 a)^2 = 10.215$ for the fundamental mode of a 'disk' resonator. Using Equation (S11), we calculate Young's modulus of β-Ga$_2$O$_3$ from the measured fundamental mode resonances. Here, we use $\rho$=5.95 g/cm$^3$ as the mass density, and Poisson's ratio of $\nu = 0.2$. By choosing 5 devices operating in the 'disk' regime with uniform, complete circular suspension (Figure S5 shows images of devices with non-ideal geometries that are not included), we extract the Young's moduli of β-Ga₂O₃ nanoflakes. An averaged Young's modulus extracted from these 5 devices is 261GPa.

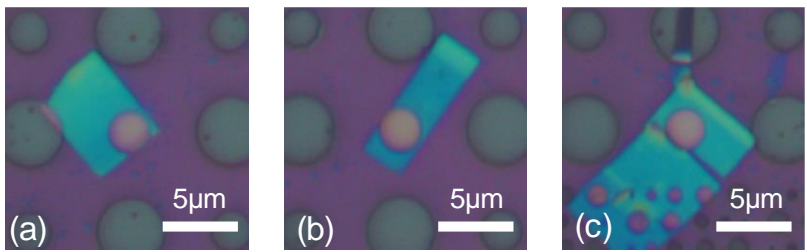

**Figure S5**. Microscopy images of β-Ga₂O₃ nanoresonators with non-ideal geometries. (a) & (b) Microtrenches are not completely covered. (c) A thickness step exists in the suspended region.

Further, we analyze the devices in the 'membrane' regime where tension dominates, where the term $\gamma a^2/D \gg (k_m a)^2$. Thus Equation (S8) becomes

$$\omega_m = \frac{(k_m a)}{a}\sqrt{\frac{\gamma}{\rho \cdot t}} \,. \tag{S12}$$

We find that in the 'membrane' regime, the resonance frequency is following the power law of $f \propto t^{-1/2}$, while in the 'disk' regime, the resonance frequency is proportional to the device thickness, $f \propto t$.

We analyze the dependency of resonance frequencies in different modes on device thickness for β-Ga₂O₃ circular drumhead resonators by plotting $\omega$-$t$ plot using Equation (S8) (Figure 3 in the Main Text). We use $E_Y = 261$ GPa which is extracted experimentally, Poisson's ratio $v = 0.2$, and radii of $r = 2.6$ μm and 1.6 μm. The plots show that 2 of the thinner devices (thicknesses of 23 nm and 42 nm) with diameter of $d \approx 5.2$ μm fall in the transition regime between the 'membrane' and 'disk', and all the other devices are in the 'disk' regime. The devices before annealing all show 'disk'-like $\omega$-$t$ behavior (Figure S6).

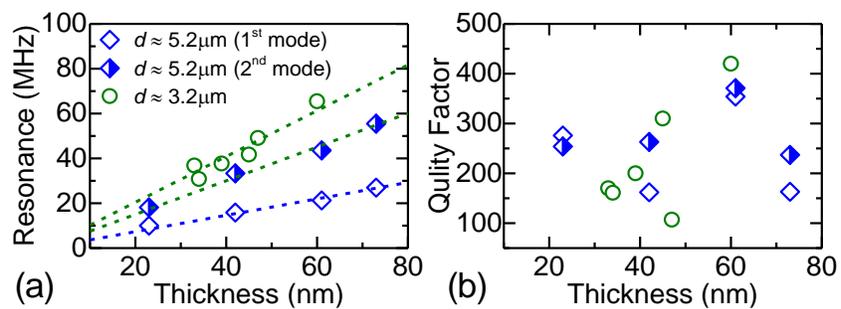

**Figure S6**. Fitted data from the measured resonances: (a) resonance frequencies and (b) quality (*Q*) factors of all devices before annealing. The hollow symbols represent the fundamental mode resonances of the devices. The half-filled symbols represent the 2$^{nd}$ modes of the devices.

## S5. Summary of Devices

Here we summarize in Table S2 the parameters and specifications of all the devices we studied in this work.

**Table S2**. List of Devices and Their Measured Parameters and Specifications

| Device # | Dia-meter $d$ (μm) | Thick-ness $t$ (nm) | Mode # $m$ | Before Annealing | | | After Annealing | | | |
|---|---|---|---|---|---|---|---|---|---|---|
| | | | | Resonance Frequency $f_m$ (MHz) | Quality Factor $Q$ | Dissipation Rate $f_m/Q$ (MHz) | Resonance Frequency $f_m$ (MHz) | Quality Factor $Q$ | Dissipation Rate $f_m/Q$ (MHz) | Young's Modulus $E_Y$ (GPa) |
| 1 | 3.20 | 33 | 1 | 36.8 | 170 | 0.216 | 41.8 | 200 | 0.209 | - |
| 2 | 3.20 | 60 | 1 | 65.5 | 420 | 0.156 | 74.9 | 566 | 0.132 | 270 |
| 3 | 3.32 | 45 | 1 | 41.7 | 310 | 0.135 | 49.4 | 556 | 0.089 | 242 |
| 4 | 3.11 | 47 | 1 | 49.1 | 107 | 0.459 | 56.0 | 160 | 0.350 | - |
| 5 | 3.20 | 39 | 1 | 37.7 | 200 | 0.189 | 47.7 | 528 | 0.090 | 259 |
| 6 | 3.65 | 34 | 1 | 30.8 | 161 | 0.191 | 40.7 | 244 | 0.167 | - |
| 7 | 5.70 | 23 | 1 | 10.0 | 147 | 0.068 | 18.1 | 826 | 0.022 | - |
| | | | 2 | 18.7 | 138 | 0.136 | 28.5 | 651 | 0.044 | - |
| | | | 3 | 20.4 | 98 | 0.208 | 33.6 | 588 | 0.057 | - |
| | | | 4 | 29.8 | 124 | 0.240 | 42.5 | 365 | 0.116 | - |
| | | | 5 | 32.1 | 101 | 0.318 | 44.8 | 596 | 0.075 | - |
| | | | 6 | 36.8 | 121 | 0.304 | 53.1 | 472 | 0.113 | - |
| 8 | 5.18 | 73 | 1 | 27.0 | 163 | 0.166 | 33.0 | 244 | 0.135 | 244 |
| | | | 2 | 55.5 | 237 | 0.234 | - | - | - | - |
| 9 | 5.24 | 61 | 1 | 21.2 | 354 | 0.060 | 39.6 | 626 | 0.063 | 292 |
| | | | 2 | 43.5 | 371 | 0.117 | - | - | - | - |
| | | | 3 | 45.4 | 295 | 0.154 | - | - | - | - |
| 10 | 5.36 | 42 | 1 | 15.9 | 162 | 0.098 | 23.5 | 395 | 0.059 | - |
| | | | 2 | 33.3 | 263 | 0.127 | 41.7 | 268 | 0.156 | - |
| | | | 3 | - | - | - | 45.7 | 505 | 0.090 | - |